\documentclass{ws-jai}
\usepackage[flushleft]{threeparttable}
\usepackage[normalem]{ulem}

\usepackage{color,ulem}
\usepackage{graphicx}
\usepackage{epstopdf}
\usepackage{epsfig}
\usepackage{url}
\definecolor{ajb}{rgb}{0.6, 0.0, 0.6}
\definecolor{gh}{rgb}{0.6, 0.8, 0.6}
\catchline{}{}{}{}{}

\begin{document}


\title{Performance Analysis Techniques for Real-time Broadband RFI Filtering System of uGMRT}

\author{Kaushal D. Buch$^{\dagger}$, Ruta Kale, Kishor D. Naik, Rahul Aragade, Mekhala Muley, Sanjay Kudale, Ajith Kumar B.}

\address{
$^1$ Digital Backend Group, Giant Metrewave Radio Telescope (GMRT), NCRA-TIFR, Pune, 410504, India \\
}

\maketitle
\corres{$^\dagger$Corresponding author}

\begin{history}
\received{(to be inserted by publisher)};
\revised{(to be inserted by publisher)};
\accepted{(to be inserted by publisher)};
\end{history}

\begin{abstract}
Electromagnetic radiation from human activities, known as man-made Radio Frequency Interference (RFI), adversely affects radio astronomy observations. In the vicinity of the Upgraded Giant Metrewave Radio Telescope (uGMRT) array, the sparking on power lines is the major cause of interference at observing frequencies less than 800 MHz. A real-time broadband RFI detection and filtering system is implemented as part of the uGMRT wideband signal processing backend to mitigate the effect of broadband RFI. Performance analysis techniques used for testing and commissioning the system for observations in the beamformer and correlator modes of the uGMRT are presented. The concept and implementation of recording simultaneous unfiltered and filtered data along with data analysis and interpretation is illustrated using an example. 
For the beamformer mode, spectrogram, single spectral channel, and its Fourier transform is used for performance analysis whereas, in the correlator mode, the cross-correlation function, closure phase, and visibilities from the simultaneously recorded unfiltered and filtered is carried out. These techniques are used for testing the performance of the broadband RFI filter and releasing it 
for uGMRT users. 


\end{abstract}


\section{Introduction}
\begin{figure}[h]
\begin{center}
 \includegraphics[scale=0.6]{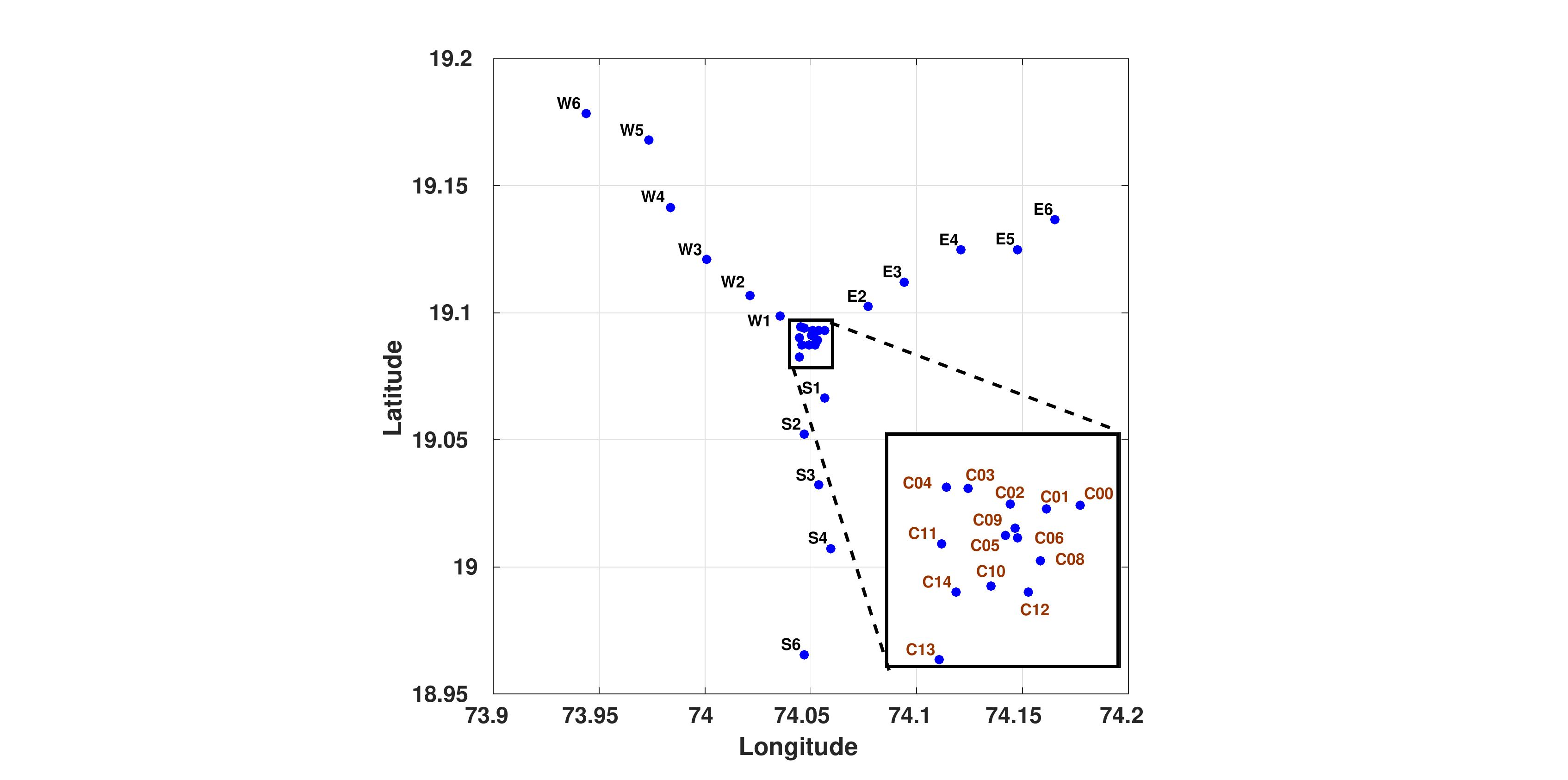}
 \advance\leftskip-3cm
\advance\rightskip-3cm
 \caption{GMRT array consisting of 30 antennas and their locations shown in Latitude (North) and Longitude (East). Central Square antennas (zoomed inset image): C00-C14 (except C07), East-arm: E2-E6, South-arm: S1-S6 (except S05), West-arm: W1-W6}
 \label{gmrtarr}
 \end{center}
\end{figure}

Giant Metrewave Radio Telescope (GMRT) \cite{swarup1991giant} is an array of 30 antennas, 14 of which are located in 1 sq. km. area and the remaining antennas are spread in East, West, and South arms as shown in Fig. \ref{gmrtarr}. The Upgraded Giant Metrewave Radio Telescope \cite{gupta2017upgraded} provides a near-seamless observing from 120 to 1450 MHz with a maximum instantaneous bandwidth of 400 MHz. Broadband RFI from powerline sparking is a major source of interference affecting the sensitivity and dynamic range of the uGMRT. 
A real-time RFI filtering system \cite{buch2019real} is implemented in the GMRT Wideband Backend (GWB) \cite{reddy2017wideband} to mitigate the broadband powerline RFI. This system computes signal statistics using robust estimation and uses threshold-based detection to identify and filter RFI \cite{buch2016towards}. The technique operates on Nyquist-sampled digital time-series from each antenna and polarization. The samples detected as RFI are replaced either by a digital noise sample or a user-defined constant value. During the system implementation phase, several performance analysis techniques were developed to understand the effects of filtering on astronomical data.

The performance of RFI filtering (excision) algorithms is generally carried out through a qualitative comparison between the cross-correlation spectrum of unfiltered and filtered data and quantitatively through improvement in the signal-to-noise ratio (SNR) \cite{fridman2001rfi, thompson2014rfi}. 
Comparing these parameters helps understand the efficacy of filtering and artefacts introduced due to the filtering process. Depending on the nature of RFI, the effects can be observed in different data products of the telescope signal processing chain. Testing of RFI mitigation can be carried out efficiently using software platforms
\cite{deller2010software} due to the ease and flexibility of implementing correlators. An example of an RFI mitigation comparison test carried out on unfiltered and filtered pulsar data by recording the raw voltage (output from analog-to-digital converter) followed by processing the signals offline is shown in \cite{ramey2019real}. Such performance checks are useful before going into the detailed analysis of the effects of filtering on astronomical imaging or time-domain astronomy.

This paper describes the performance analysis techniques developed and used for testing the real-time broadband RFI filtering system of the uGMRT. Since the filtering system implementation has gone through several phases, the analysis techniques have evolved over a period of time.  Several new techniques and parameters were introduced over the initial testing scheme \cite{buch2018implementing} to qualify the filtering system and study the effects on the beamformer and correlator modes of the GWB. These techniques have been used over for three years on various uGMRT observations. They have helped in identifying the problems, fine-tuning the filter configuration, and releasing the system for astronomical observations \footnote{http://www.ncra.tifr.res.in/ncra/gmrt/gmrt-users/online-rfi-filtering}. The basis of these techniques is the simultaneous comparison between the various parameters of unfiltered and filtered data. 

The paper is organized as follows: Section 2 provides an overview of the broadband RFI at GMRT and the implementation of the RFI filtering system; Section 3 provides the test setup for simultaneous testing; Section 4 provides the uGMRT system setup for example observation; Sections 5 and 6 provide the performance analysis in the beamformer and correlator modes respectively and Section 7 provides a discussion on simultaneous testing and effects related to RFI filtering.

\section{Broadband RFI at GMRT and real-time RFI filtering system}
Power-line RFI manifests itself as a collection of impulsive events in the time domain. 
However, the behavior as seen at the input of the digital backend system can vary depending on the source of RFI, propagation characteristics, and the response of the receiver system to these impulsive events. 
The potential sources of powerline interference are spread across the GMRT array in terms of sparking on high-power transmission lines, faults on single-phase and three-phase transformers and domestic power distribution lines \cite{swarup2008power, raybole2010external}. As a typical example of the potential sources, there are about 115 known transformer installations distributed in $\sim$ 4 km radius from the centre of the array (with C02 antenna as reference). There are also high-power transmission lines near the outermost arm antennas of the GMRT array \cite{raybole2010external}.

\begin{figure}[h]
\begin{center}
 \includegraphics[scale=0.5]{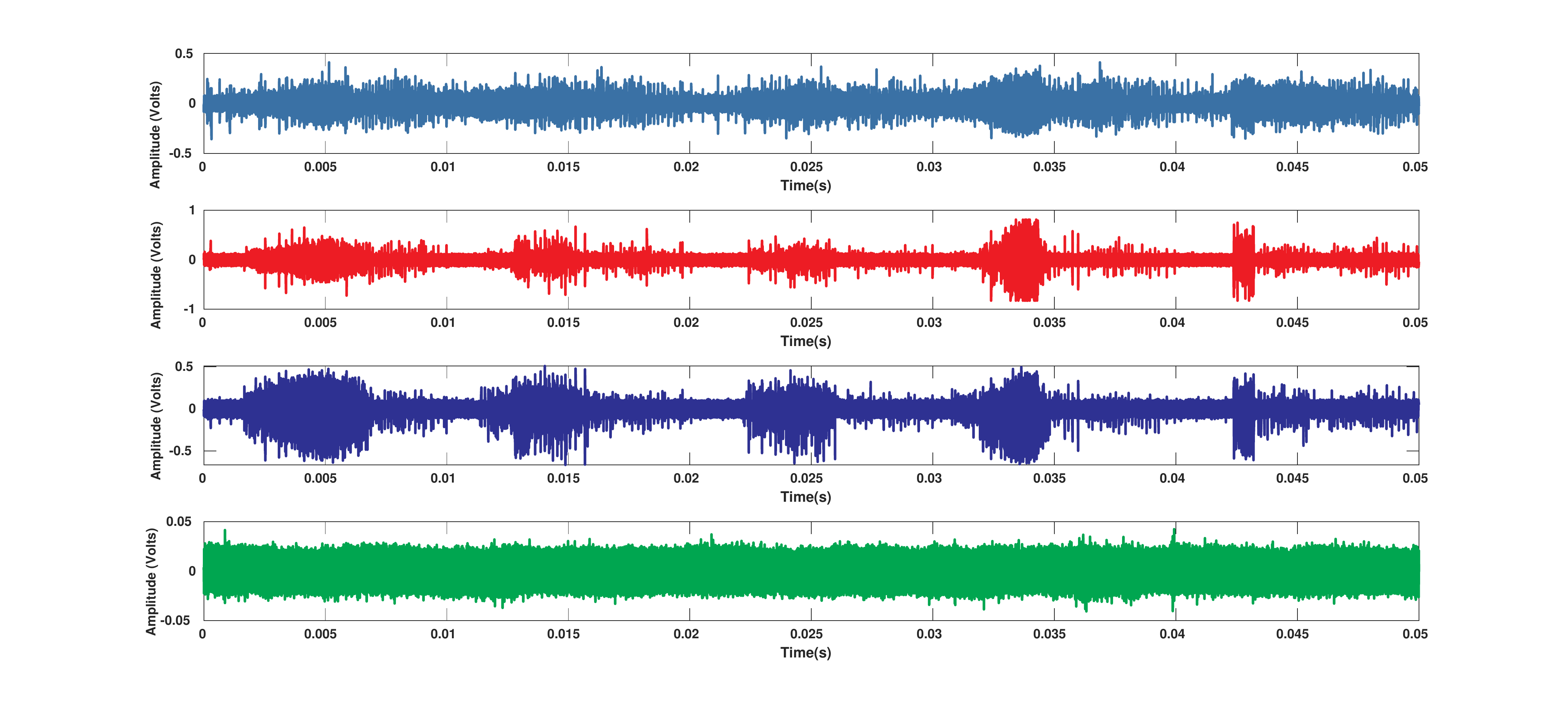}
 \advance\leftskip-3cm
\advance\rightskip-3cm
 \caption{50ms plot of voltage data acquired simultaneously using a 4-channel oscilloscope at 5$ns$ time sampling. First subplot from top is Band-2 (C01 antenna), second subplot is Band-3 (C05 antenna), third subplot is Band-4 (C06 antenna) and fourth subplot is Band-5 (C08 antenna). Voltage values for Band-2 to 5 are 100mv/div, 200mV/div, 200mV/div and 20mV/div respectively.}
 \label{rfimso}
 \end{center}
\end{figure}

Powerline RFI is the strongest in uGMRT Band-2 (120-250 MHz) and is observed in Band-3 (250-500 MHz) and Band-4 (550-850 MHz). A time-domain plot acquired simultaneously from four nearby GMRT antennas, each observing in different uGMRT bands, using an oscilloscope connected at the input of the digital backend system is shown in Fig. \ref{rfimso}. As is seen in this plot, the strength of the RFI changes across uGMRT bands. The amplitude and distribution of impulsive events may vary between the bunches \cite{swarup2008power, loftness1997power}. The periodicity is typically 10ms or 20ms (due to 50 Hz powerline frequency) or less depending on the fault on the powerline \cite{swarup2008power}. The strength of powerline RFI, particularly those occurring due to gap discharge, is observed up to 1 GHz \cite{maruvada2000corona}. Hence, it affects observations in bands 2, 3, and 4 of uGMRT.

The strength of powerline RFI at GMRT is typically 10-20 dB more than the system noise \cite{swarup2008power} which causes a reduction in SNR and dynamic range of the receiver. Hence, a real-time RFI mitigation system is developed to selectively filter impulsive RFI. The system consists of a robust statistical estimator to determine the filtering threshold of the input signal. The current implementation uses a Median-of-MAD (MoM) estimator, a variant of Median Absolute Deviation (MAD). A detector follows the estimator block to discriminate RFI from the astronomical data. The detected RFI samples are replaced by digital noise samples or user-defined constant values. The filtering process is carried out on Nyquist-sampled digital time-series for each antenna and polarization\cite{buch2019real}.

The RFI filtering system is an FPGA-based implementation on a ROACH-1 board\footnote{https://casper.ssl.berkeley.edu/wiki/ROACH}from CASPER (Collaboration for Astronomy Signal Processing and Electronics Research) \cite{hickish2016decade}. It can process four inputs digitized to 8-bit precision at 800 MHz sampling frequency. Currently, the GWB connects four different antennas of a particular polarization to a single ROACH-1 board. Each input goes through an RFI filtering system, and the filtered data is sent to the CPU/GPU cluster through the 10 Gigabit Ethernet ports. The GWB operates in two main signal processing modes – correlator and beamformer \cite{reddy2017wideband}.

\section{GWB System Setup for Simultaneous testing}

Performance analysis of an RFI filtering technique is difficult to assess in the normal mode of the GWB operation where the filtering is applied to all the inputs. This default mode does not provide an unfiltered and a filtered copy simultaneously for comparison. Hence, for a fair comparison, the analysis needs to be carried out in a mode which allows simultaneous processing and recording of antenna inputs with and without the filter. For implementing this, the 30-antenna 
dual-polarization GWB system is programmed to simultaneously provide unfiltered and filtered outputs. The two main steps for the setup are to digitally copy the data from one antenna to the other (inside the FPGA) and to program the correlator such that it treats the original antenna (unfiltered) and its digital copies (filtered) as a single antenna. The latter is achieved in the GWB by making antenna coordinates and hence the delay correction identical for the original antenna and its digital copy. The advantage of the digital copy scheme is that it allows for a fair comparison between the unfiltered and filtered data.

\subsection{1:2 Digital Copy Mode} \label{12digcpy}

\begin{figure}[h]
\begin{center}
 \includegraphics[scale=0.4]{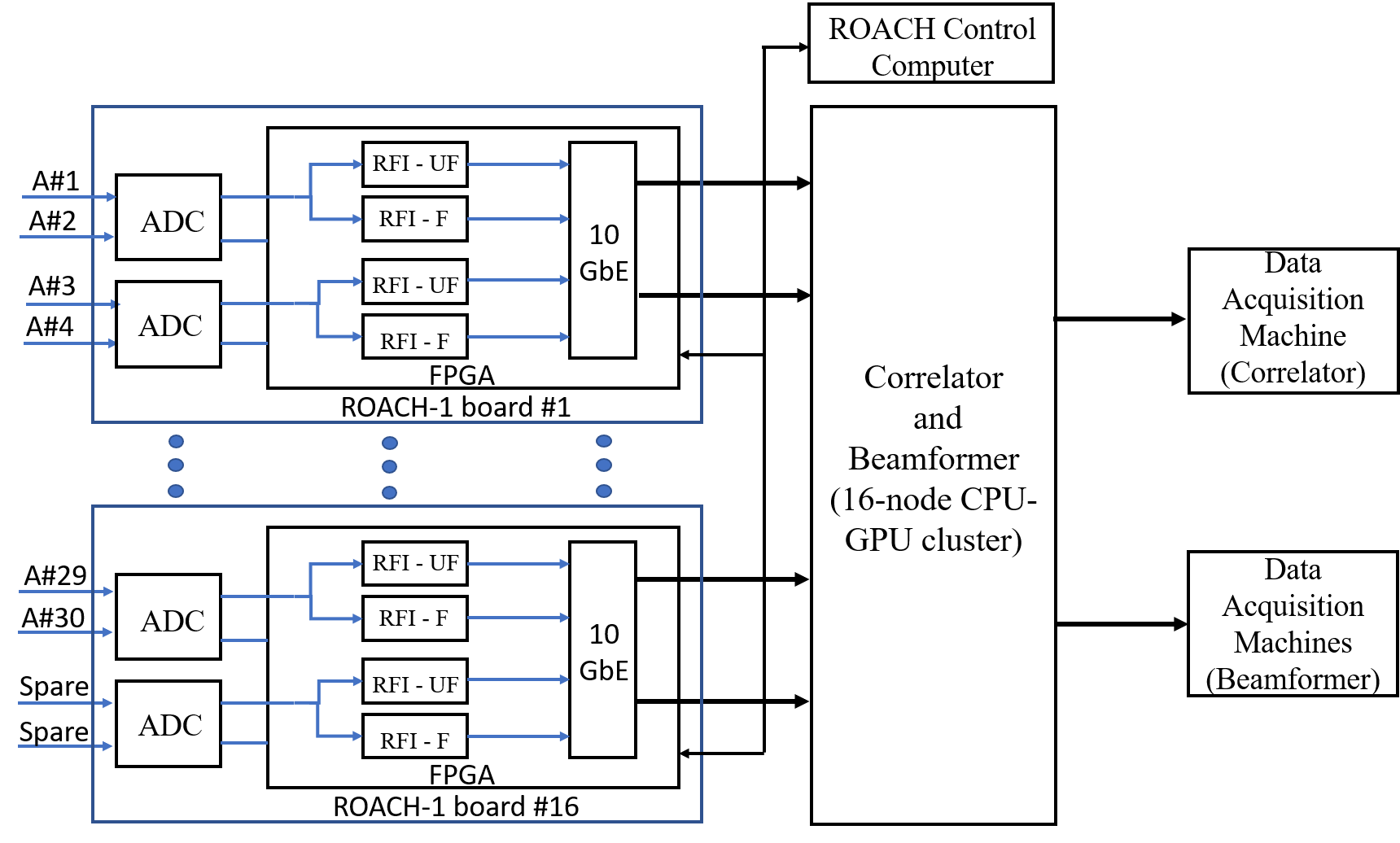}
 \advance\leftskip-3cm
\advance\rightskip-3cm
 \caption{Block diagram of GWB setup for the 1:2 digital copy mode.}
 \label{rfibd1}
 \end{center}
\end{figure}

In this mode, the data from one antenna is digitally copied into the next antenna on the FPGA. Fig. \ref{rfibd1} shows the location of the RFI filtering block, the mode configuration, and the overall arrangement for simultaneous testing. Table \ref{anttable} provides the configuration in which the sequential antenna inputs to the GWB are connected in the simultaneous testing mode. For example, on a single ROACH-1 board, C00 is copied to C01 and C02 is copied to C03, and so on. Fifteen antennas that are digitally copied undergo RFI filtering. 

\begin{table}[h]
\caption{GMRT antenna configuration for the 1:2 digital copy mode} 
\label{anttable}
\vspace{0.2cm}
\resizebox{\textwidth}{!}{%
\begin{tabular}{|c|c|c|c|c|c|c|c|c|c|c|c|c|c|c|c|c|c|c|c|c|}
\hline
 Unfiltered& C00& C02 &C04 &C06 &C09 &C11 &C13 &E02 &E04 &E06 &S02 &S04 &W01 &W03 &W05\\
\hline
Filtered& C01 &   C03 &C05 &C08 &C10 &C12 &C14 &E03 &E05 &S01 &S03 &S06 &W02 &W04 &W06\\
\hline
\end{tabular}}
\end{table}

\subsection{1:4 Digital Copy Mode}

\begin{figure}[h]
\begin{center}
 \includegraphics[scale=0.4]{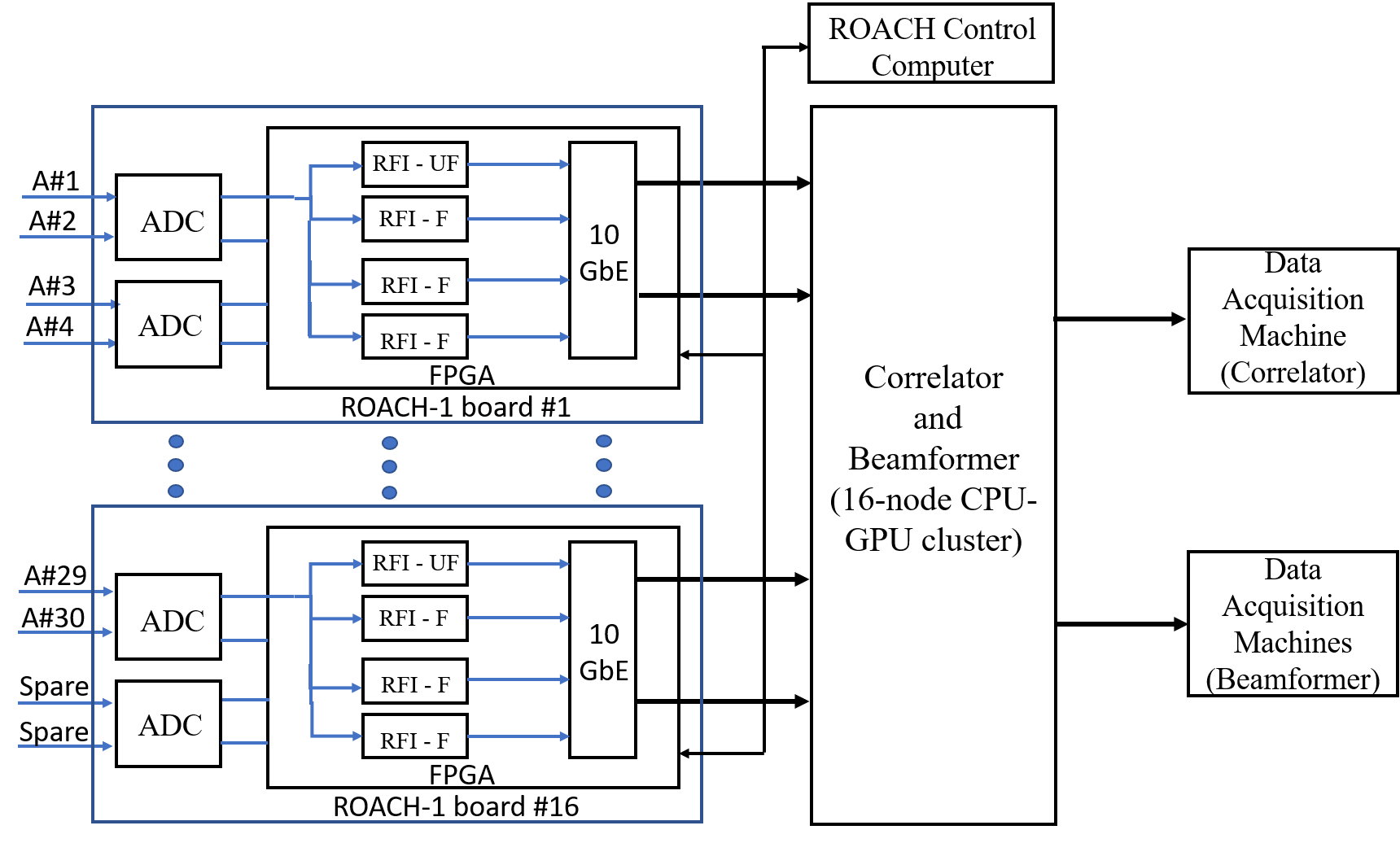}
 \advance\leftskip-3cm
\advance\rightskip-3cm
 \caption{Block diagram of GWB setup for the 1:4 digital copy mode}
 \label{rfibd2}
 \end{center}
\end{figure}

In this mode, the input data from one antenna is digitally copied to three other antenna inputs, which for the GWB system, is the maximum number of copies possible. This configuration compares the different filtering algorithms or a particular algorithm with different threshold or replacement values for the RFI system testing. Fig. \ref{rfibd2} shows the location of the RFI filtering block, the digital copy configuration, and the overall arrangement for simultaneous testing using this mode. The delay inside the GWB system is corrected to provide the same delay to the other three inputs, i.e. C00 antenna’s delay is applied to its digital copy, which is C01, C02, and C03 antennas. Table \ref{anttable14} shows the configuration of the input and the digital copies. Other possible configurations in 1:4 mode are C03 antenna data copied to C00, C01, C02, and similarly for the other antennas. The choice to select input antenna and the copies can be modified on the baselines required for the test.

\begin{table}[h]
\caption{GMRT antenna configuration for the 1:4 digital copy mode} 
\label{anttable14}
\vspace{0.2cm}
\resizebox{\textwidth}{!}{%
\begin{tabular}{|c|c|c|c|c|c|c|c|c|}
\hline
 Unfiltered& C00& C04 &C09 &C13 &E04 &S02 &W01 &W05 \\
\hline
Filtered-1& C01 & C05 &C10 &C14 &E05 &S03 &W02 &W06\\
\hline
Filtered-2& C02 & C06 &C11 &E02 &E06 &S04 &W03 &\\
\hline
Filtered-3& C03 & C08 &C12 &E03 &S01 &S06 &W04 &\\
\hline
\end{tabular}}
\end{table}

\section{uGMRT observation: RFI Test Setup and Preprocessing}

The test setup, GWB configuration, and data pre-processing are explained through a test observation carried out in Band-4 (550-850 MHz) during the monsoon season (when the powerline RFI is at its peak). 
Another reason for choosing this observing band is that it has lesser narrowband RFI (from radio communication transmitters) as compared to the other uGMRT bands. This helps in studying the effects of broadband RFI filtering in isolation.

For testing the RFI filtering system, observing an unresolved calibrator source is preferred. The cross-correlation value across the baselines for an unresolved radio source is the same. Hence, any changes occurring due to RFI can be easily identified.

The observation and system settings for the test observation are mentioned in Table 3. 


\begin{table}[h]
\label{conftable}
\caption{Observation and System Settings for the example RFI test observation}
\centering
\begin{tabular}{@{}lll@{}} \toprule
Parameters & Value \\ \colrule
Source & 3C48 \\
Observation date & 14 June 2018 \\
Observing duration & 20 min. \\
No. of antennas & 9 \\
RF Band & Band 4 (550-850 MHz)\\
Processing Bandwidth & 200 MHz \\
Input RF Bandwidth & 100 MHz (550-650 MHz) \\
ADC quantization & 8 bits \\
Number of Spectral Channels & 2048 \\
Spectral resolution & 97.65 kHz \\
RF Band Start Frequency & 550 MHz \\
Frequency for channel number 700 & 618 MHz \\
Integration Time (Correlator) & 0.671s \\
Integration Time (Beamformer) & 81.92 $\mu$s \\
Mode (Correlator) & Total Intensity \\
Mode (Beamformer) & Total Intensity \\
Data length (Correlator) & 1813 samples \\
Data length (Beamformer) &   13328384 samples   \\
Antennas for Correlator &  C00,C02,C04,C06,C09, C11,C13,E04,E06 \\
Antenna for Beamformer &    C09 \\
Filtering mode & 1:2 digital copy \\
RFI Filter Threshold & 3$\sigma$ \\
Replacement Option & Digital Noise \\
Average Percentage RFI (per antenna) & 2-3\% \\
 \botrule
\end{tabular}
\end{table}

Before recording the data, the input Radio Frequency (RF) power from all the antennas is equalized. The general rule followed for uGMRT observations is to have sufficient input power-level to the Analog-to-Digital (ADC) enough to occupy up to 6 bits of an 8-bit quantized signal to have enough headroom for the addition of RFI. The detection performance depends on the power level of system noise.

The simultaneous recording has been carried out for the beamformer and correlator modes at the highest time resolution provided by the GWB system.



\subsection{Data Extraction}

Typically, the powerline RFI observed at GMRT is broadband and unpolarized. Hence the example test results shown in the subsequent sections (except for Section 5.1) are for the single spectral channel data. The data are extracted from the recorded beamformer and correlator (visibility) outputs. In general, a spectral channel away from strong narrowband interference and near the centre of the band is chosen. Here, the results from spectral channel number 700 (corresponding to 618 MHz RF) are presented.


Time-frequency and single spectral channel beamformer data are extracted for a desired number of timestamps. The extracted files are converted to a decimal format for ease of visualization and processing. The timestamp extracted from the header files is used to indicate the exact observation start time. 

Correlator data are recorded as real and imaginary values for each spectral channel, baseline, and polarization in the LTA (long-term accumulation) file. LTA is a GMRT-specific file format, and the desired baseline's amplitude and phase values in decimal format are extracted using the 'xtract' \cite{bhatnagar1997xtract} utility developed at GMRT. xtract is also used to extract single or multiple spectral channel output for a desired number of timestamps.

\section{Performance Analysis in the Beamformer Mode}

 A single antenna input (unfiltered) and its digital copy (filtered) are used to compare the SNR, the effect on 50 Hz powerline frequency, and its harmonics. Qualitatively, the analysis can be carried out on power spectrum of the beamformer output by observing the effect on the spectrogram and single spectral channel.

\subsection{Spectrogram}

\begin{figure}[h]
\begin{center}

         \includegraphics[trim={0cm 0cm 0cm 0cm}, clip,height=4.0cm,angle=0]{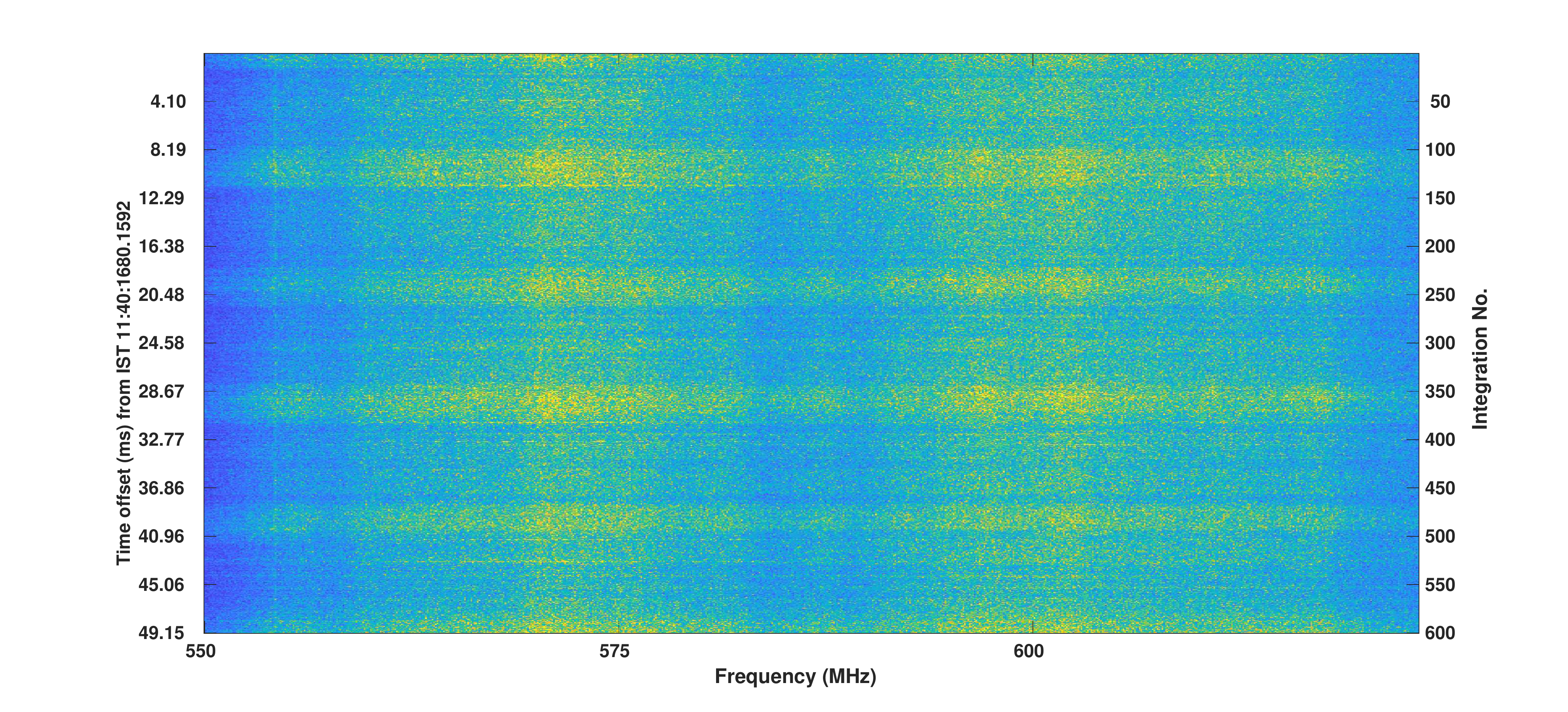}
        \includegraphics[trim={0cm 0cm 0cm 0cm}, clip,height=4.0cm,angle=0]{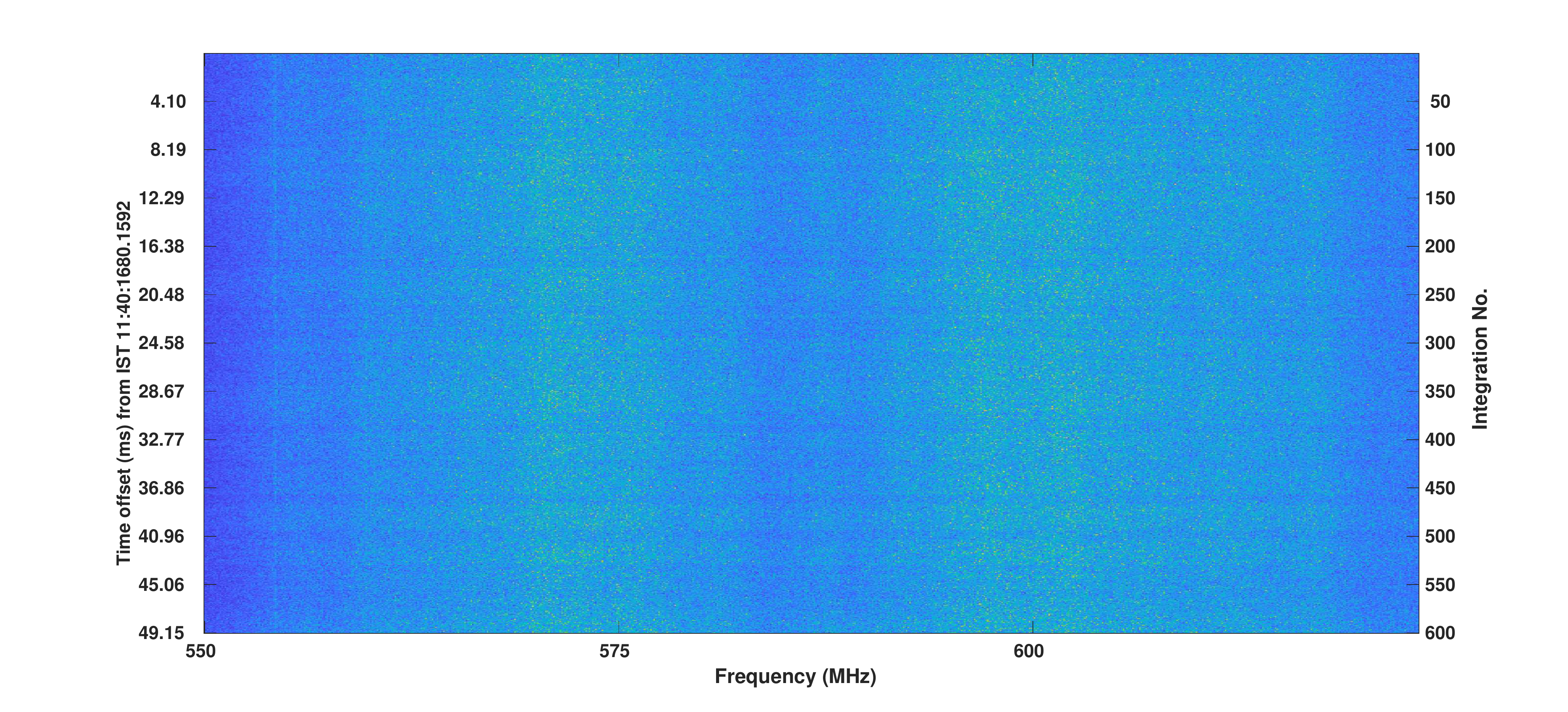}
          \caption{Spectrogram - Unfiltered (left) and Filtered (right).  The Y-axis shows time offset from an instance of time period (in Indian Standard Time IST) during the observation. Time value on Y-axis is the number of spectra times the integration time. X-axis shows the RF corresponding to each spectral channel in the range of 550-650 MHz. The Y-axis on the right shows the integration number. Different colors in the plot show power (in arbitrary units) and ranges from 10 (blue) to 4500 (yellow).}
 \label{beamspec}
 \end{center}
\end{figure}

A short-duration power spectrum for unfiltered and filtered beamformer output shows powerline RFI as a periodic increase in the power across the entire spectrum. Fig. \ref{beamspec} shows the spectrogram of an unfiltered and filtered beamformer output of a GMRT antenna. An increased broadband power for a duration of approximately a millisecond occurring with a periodicity of $\sim$10 ms corresponds to the half-cycle period of the powerline frequency.\footnote{Powerline frequency is 50 Hz in India} .

The filtered spectrogram shows suppression of periodic increase in broadband power, indicating that broadband RFI has been detected and filtered (replaced with digital noise samples).

\begin{figure}[h]
\begin{center}
 \includegraphics[scale=0.4]{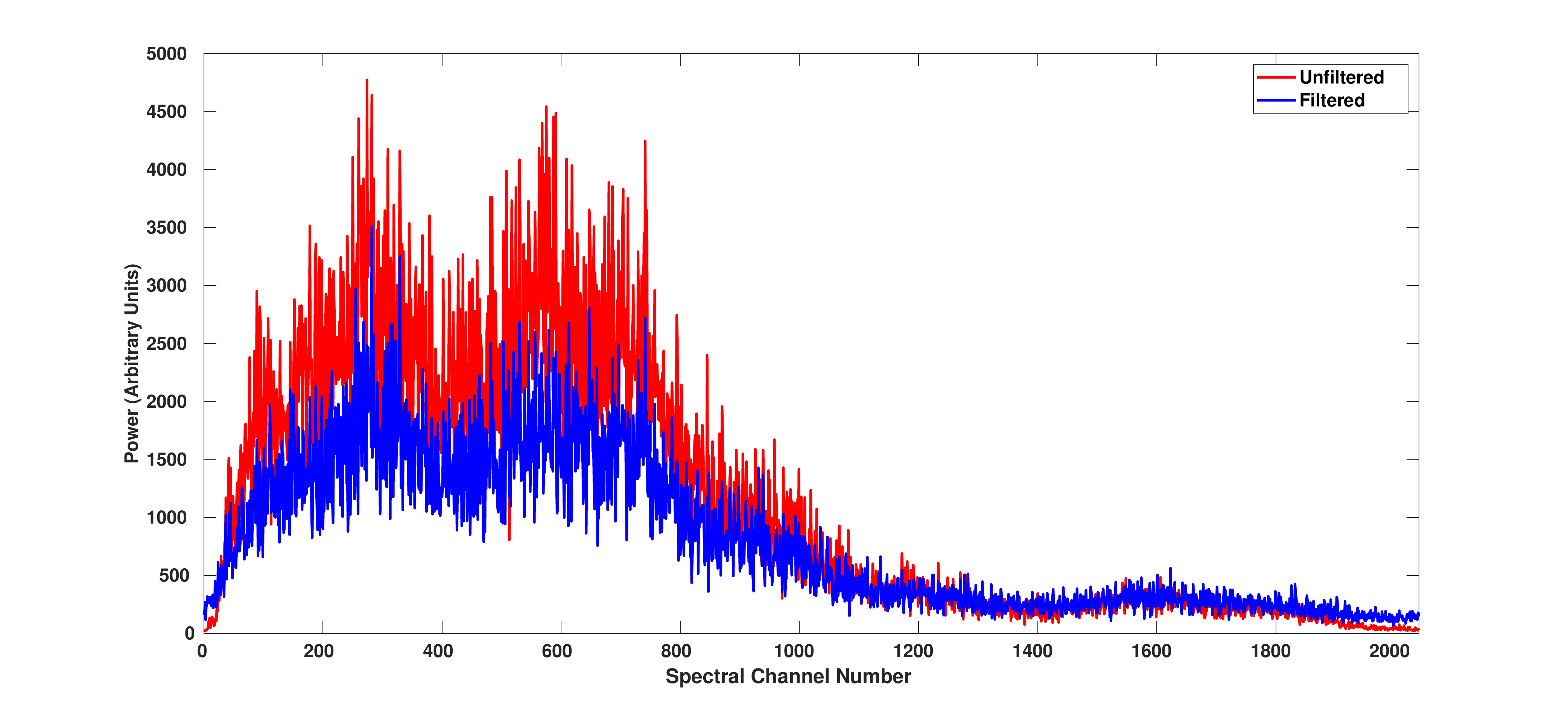}
 \advance\leftskip-3cm
\advance\rightskip-3cm
 \caption{The unfiltered (red) and filtered (blue) power spectra are shown. The reduction in the scatter in the filtered spectrum is the effect of broadband RFI filtering. The shape of the individual spectra correspond to the spectral response of the receiver.}
 \label{beamband}
 \end{center}
\end{figure}

The effect is also analyzed on a single spectrum extracted from the spectrogram shown in Fig. \ref{beamspec}. A comparison spectrum for integration number 120  is shown in Fig. \ref{beamband}. The spectrum for this particular time instance is chosen as it corresponds to the presence of powerline RFI. The increased broadband power due to RFI, as seen in the unfiltered case, is suppressed in the filtered counterpart.

A quantitative analysis of the improvement will be described in the subsequent sections.

\subsection{Single Spectral Channel}

\begin{figure}[h]
\begin{center}
 \includegraphics[width=\textwidth]{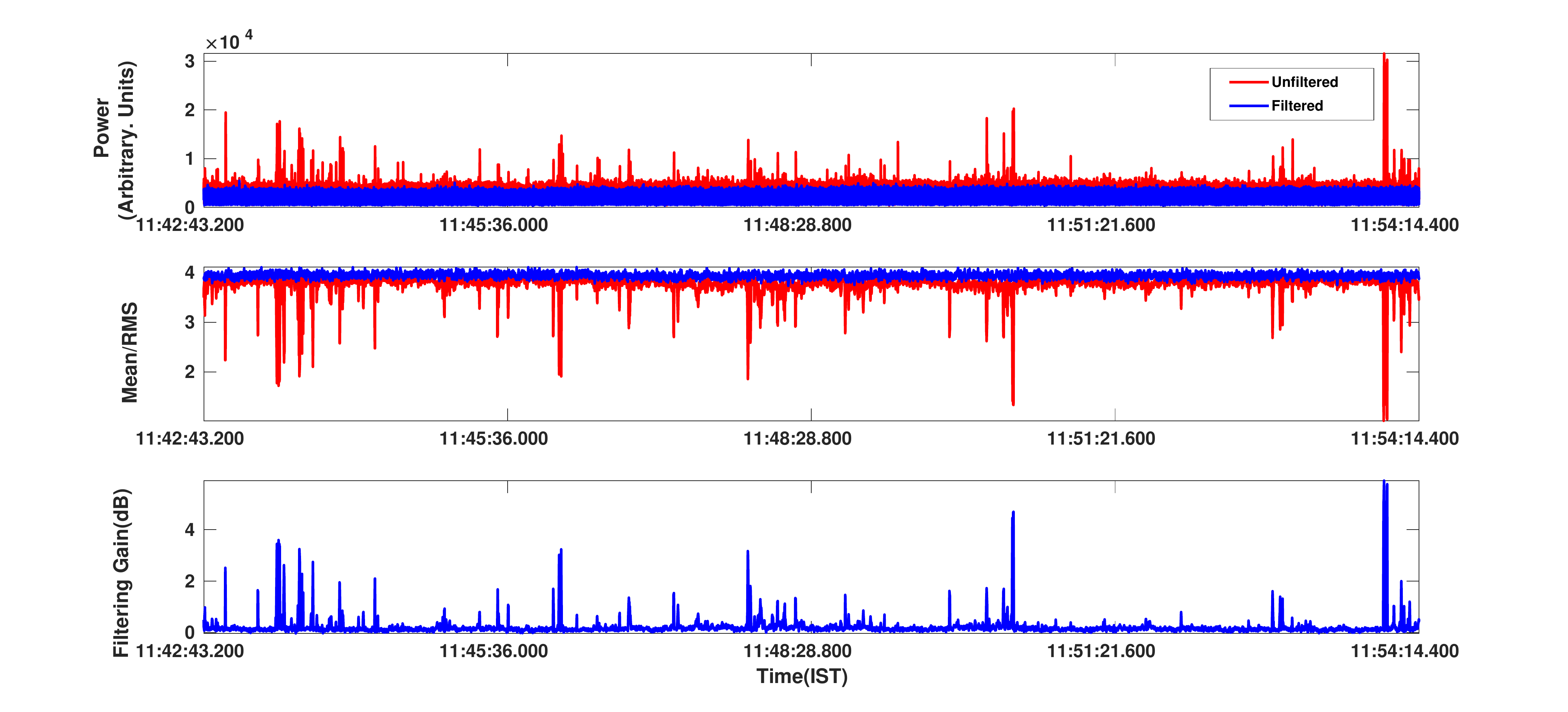}
 \advance\leftskip-3cm
\advance\rightskip-3cm
 \caption{Comparison of power for a single spectral channel (top plot), SNR (middle plot) and filtering gain (bottom plot) versus time.}
 \label{beamts}
 \end{center}
\end{figure}

A single spectral channel can be used to observe the effect of broadband RFI. A comparative analysis of its filtered counterpart can be used for analyzing the improvement in SNR. The first subplot of Fig. \ref{beamts} shows beamformer output power (arbitrary units) plot versus time (IST). The increase in average power due to RFI in the unfiltered data is corrected in the case of filtered data which can be quantified through improvement in the SNR.

Theoretically, the SNR is represented as the ratio of mean to root mean square (RMS) of a signal and is expressed as $\sqrt{B\tau}$, where B is the bandwidth, and $\tau$ is the integration time. For this test observation, bandwidth (B) for a single spectral channel is 97.5 kHz. For the total intensity mode (both polarizations are added) used during this observation, the theoretical mean to RMS ratio would be $\sqrt{2B\tau}$ = $\sqrt{2*97.5 \rm{kHz}* 81.92\mu s}$ = 4. The SNR reduces in the presence of RFI, which is the case for unfiltered data in the second subplot in Fig. \ref{beamts}. Post-filtering, the SNR reaches its theoretical maximum value.

The processing (filtering) gain \cite{fridman2001rfi} can be calculated as the ratio of SNR (Filtered) to SNR (Unfiltered). The third subplot in Fig. \ref{beamts} shows the filtering gain for a single spectral channel. A high processing gain is observed where the SNR has improved as a result of the filtering.


 \subsection{Fourier Transform of Single Spectral Channel}
 
 Fourier Transform of a single spectral channel is used to find the periodicity in the occurrence of RFI. As mentioned earlier, the powerline RFI occurs at multiples of the powerline frequency. Hence, the magnitude spectrum of a single spectral channel should show a presence of 50 Hz, 100 Hz, and their harmonics. 
 
\begin{figure}[h]
\begin{center}
 \includegraphics[width=\textwidth]{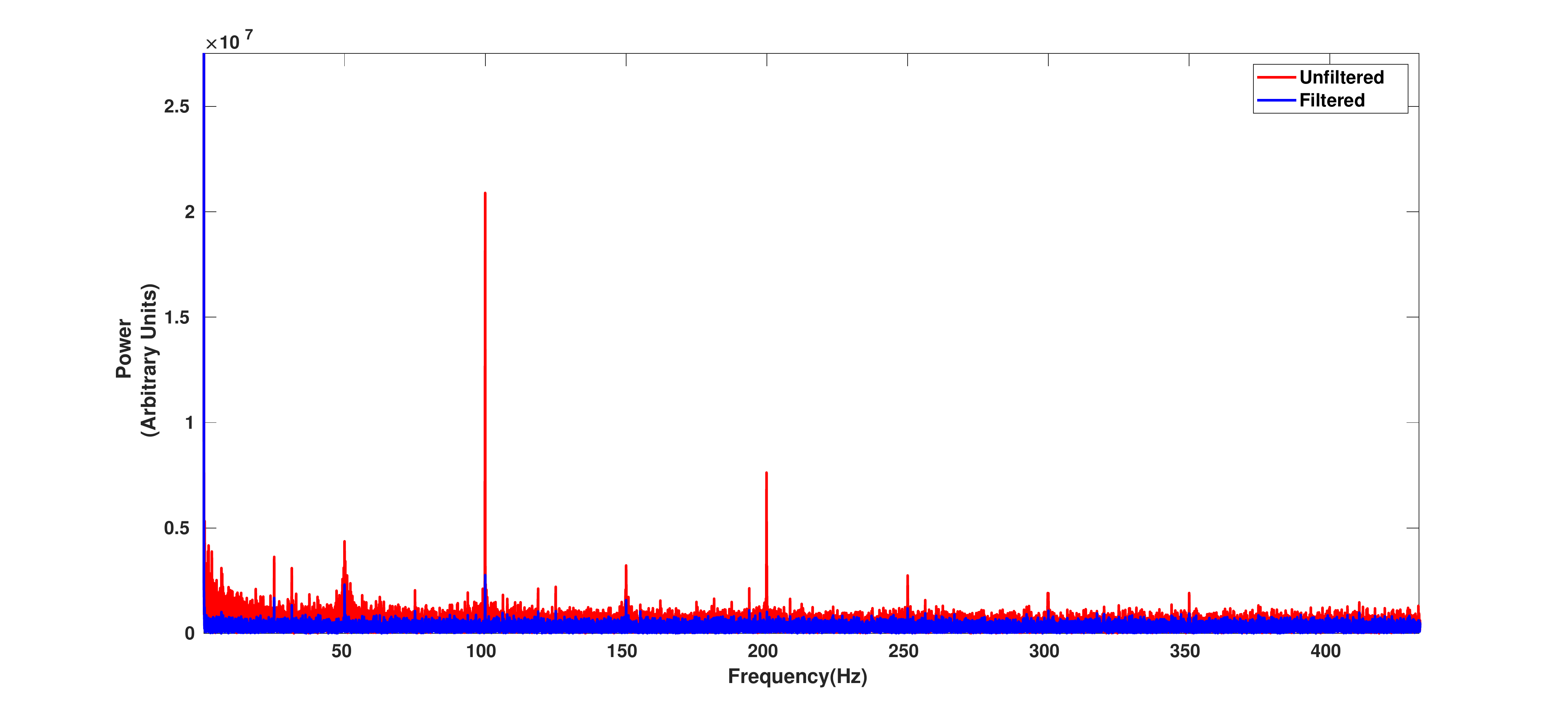}
 \advance\leftskip-3cm
\advance\rightskip-3cm
 \caption{Comparison plot of magnitude spectrum of single spectral channel showing the suppression in the powerline frequency and harmonics.}
 \label{beamfft}
 \end{center}
\end{figure}

A Fast Fourier Transform (FFT) operation is carried out on single spectral channel data. To achieve better frequency resolution, the number of the FFT points is the same as the length of the data. A comparison between unfiltered and filtered data shows the suppression in the fundamental powerline frequency component and harmonics. Fig. \ref {beamfft} shows the magnitude spectrum of single spectral channel data (shown in earlier subsection). The frequency shown on X-axis is derived from the integration time and spectral channel resolution. The rejection at 100Hz is the ratio of unfiltered to filtered magnitude converted to dB scale, which is 8.7 dB. The dominance of 100 Hz harmonic has been reduced significantly.
 
 \section{Performance Analysis in the Correlator Mode}
 
 Correlator data are recorded at a time resolution of 0.671 s, which is the finest time resolution available in GWB. The cross-correlation amplitude and phase for all the available baselines (antennas pairs) are used for the analysis methods described here. 
 
 Correlator data analysis helps understand the effects of filtering on the cross-correlation measurement (visibilities) and spatial correlation of RFI over baselines. This section covers the effects of filtering on the cross-correlation function and closure phase through the example test observation.

\subsection{Cross-correlation Function}

\begin{figure}[h]
\begin{center}
 \includegraphics[scale=0.40]{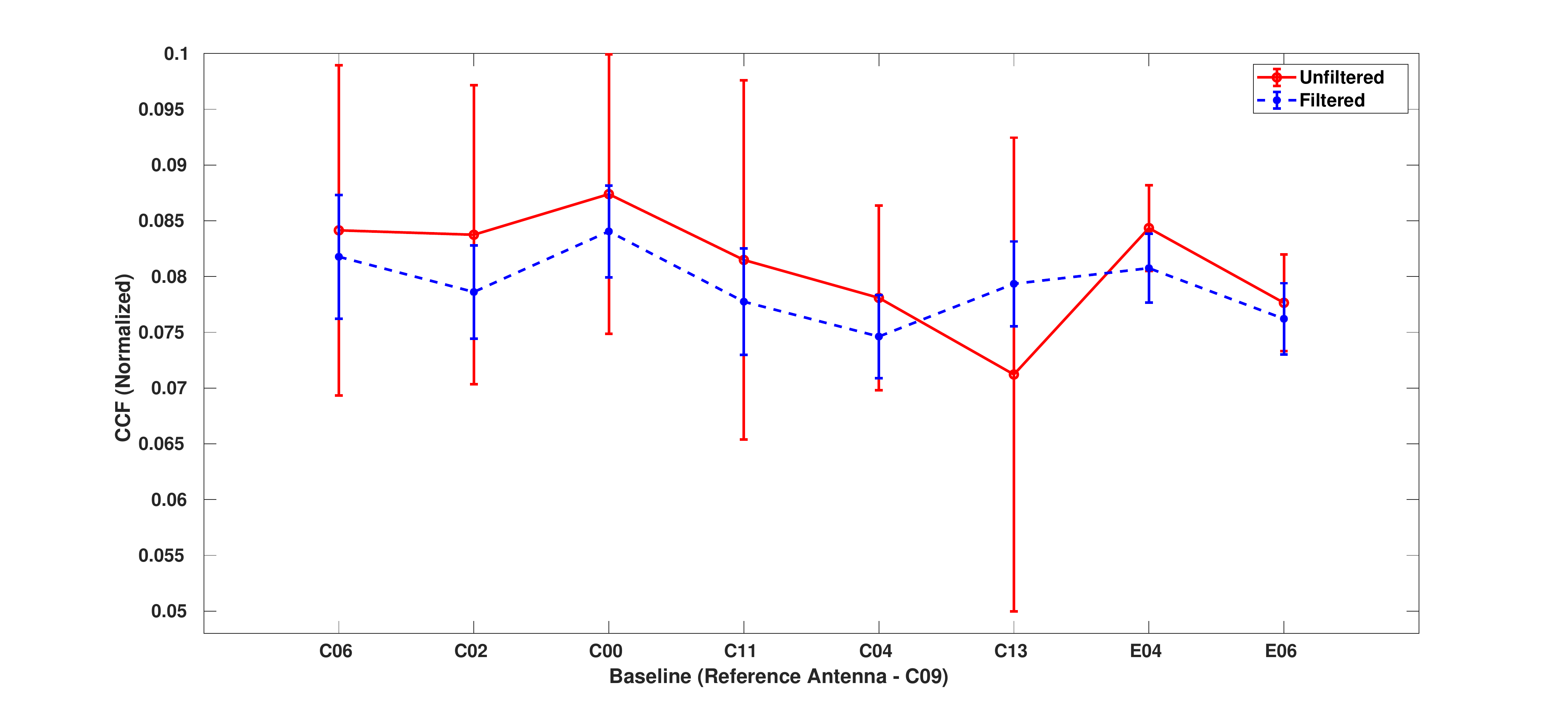}
 \advance\leftskip-3cm
\advance\rightskip-3cm
 \caption{Comparison plot for CCF of different baselines (with respect to C09 antenna. The baselines in the plot are in ascending order of baseline length for spectral channel 700 which is 103,164,660,669,674,1224,8390 and 13016 metres respectively.)}
 \label{corrccf}
 \end{center}
\end{figure}

Powerline RFI at GMRT is spatially localized and affects the cross-correlation measurement for co-located antennas, particularly those located in the central square of the array. Fig. \ref{corrccf} shows cross-correlation function (normalized) for different baselines with the C09 antenna as reference. Correlation information for a single spectral channel across different baselines for the entire duration of the observation is extracted. The mean and standard deviation of the cross-correlation function (CCF) are computed on data taken over the entire observing duration.

In an observation without RFI, the mean and standard deviation of CCF across the baselines remain the same for an unresolved source. In this case, the unfiltered data show more significant error bar as compared to the filtered data. A larger error bar is due to the non-random nature of RFI, which leads to a heavy-tailed distribution. Such data do not follow the radiometer equation upon integration.


For the filtered version, the error bar is reduced for all the RFI-affected baselines. The long baselines are not affected by RFI. The mean CCF has also been reduced for the filtered data. The reduction is because of the high amplitude RFI being replaced by digital noise (with a statistical distribution similar to the parent distribution) which causes a reduction in the average value of the correlation power spectrum. In terms of quantitative improvement, the reduction in the error bar can be represented as the ratio of the standard deviation of unfiltered to filtered. For this example, the best case improvement is by a factor of 5.5.


\subsection{Closure Phase}

The closure phase ($\phi$) refers to the relationship between the cross-correlation phase of three or more antennas of a radio interferometer array. This quantity allows the measurement of the cross-correlation of the source visibility and is independent of instrumental calibration errors. 
The cross-correlation phase of three antennas forms a closure triangle, and $\phi_{12} + \phi_{23} + \phi_{31}$ should be zero for a signal without RFI. In the presence of RFI, the closure phase variations depart from zero \citep[e. g.][]{kocz2015digital}. 

For testing the RFI filtering system, we form a closure triangle from three short-spacing antennas of GMRT (C00, C02, and C04) and their filtered copies (C01, C03 and C05) using the 1:2 digital copy mode (Sec.~\ref{12digcpy}). 
 Hence, a total of six baselines would be available (C00-C02 and its filtered version C01-C03, and so on).

\begin{figure}[h]
\begin{center}
 \includegraphics[scale=0.5]{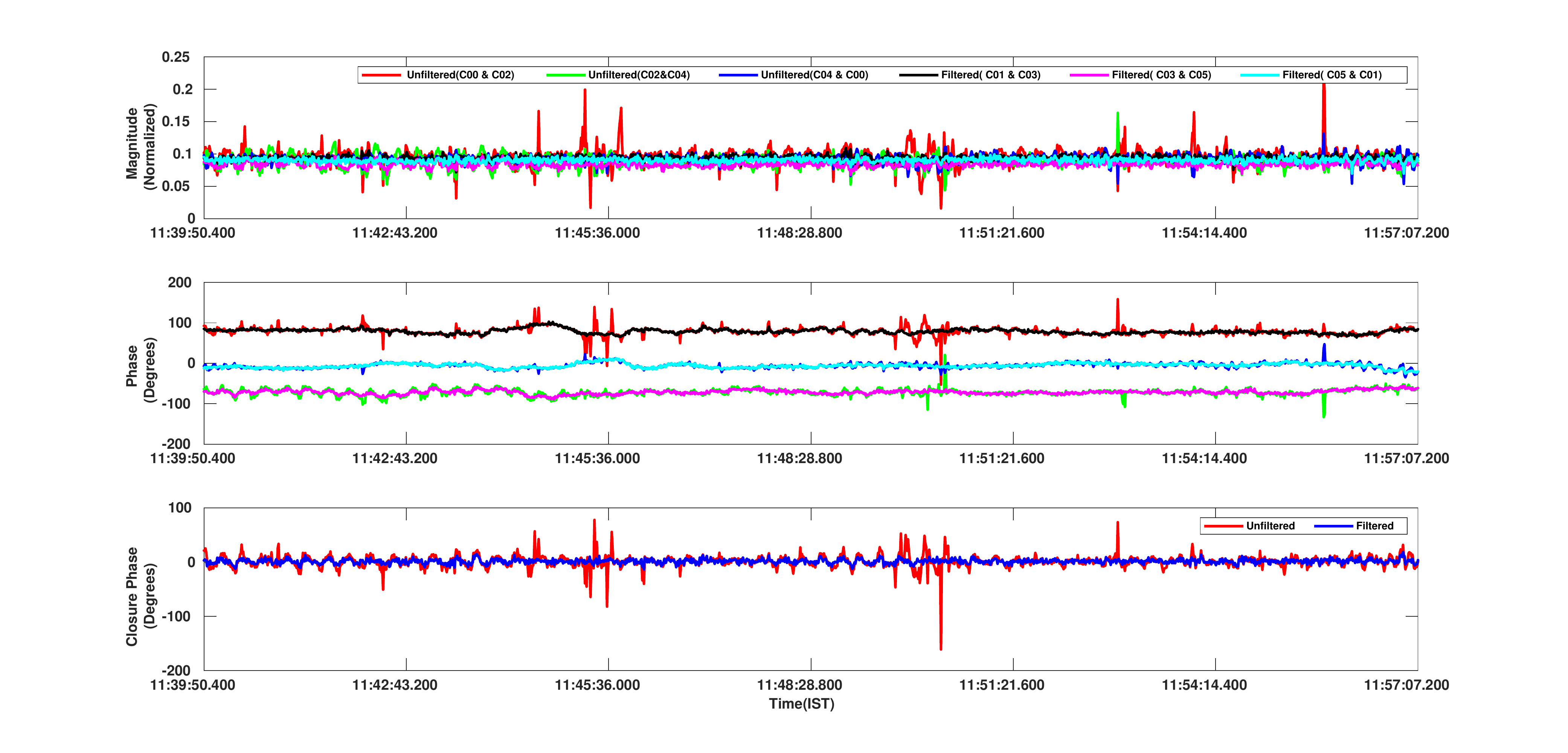}
 \advance\leftskip-3cm
\advance\rightskip-3cm
 \caption{Comparison of closure phase formed by C00,C02, and C04 antennas and their filtered counterparts. The plots shown are for a single spectral channel versus time. The baselines lengths for C00-C02, C02-C04, and C00-C04 corresponding to spectral channel number 700 are 688m, 581m, and 1263m, respectively. The Y-axis of the upper subplot shows normalized cross-correlation magnitude for the unfiltered and filtered baselines and the middle subplot shows the phase of the cross-correlation. The lower subplot shows the closure phase for the unfiltered and filtered antennas.}
 \label{corrcp}
 \end{center}
\end{figure}

The upper subplot of Fig. \ref{corrcp} shows the normalized cross-correlation amplitude for the unfiltered and filtered baselines. Strong RFI is observed intermittently in the cross-correlation amplitude, which is filtered in the counterpart. In the middle subplot, the cross-correlation phase of unfiltered baselines shows fluctuations due to RFI, which are removed in the filtered baselines. The lower subplot is the closure phase which shows variations in the range of $\pm{180}$ degrees. For the unfiltered data, the variation is in the range of -161 to +78 degrees. In contrast, the post-filtering closure phase is in the range of $\pm{15}$ degrees indicating the large fraction of RFI has been filtered, and the closure phase is close to zero degrees. 

\subsection{Calibrated visibility amplitudes}

\begin{figure}[h]
\begin{center}
        \includegraphics[trim={1cm 4cm 1cm 3cm}, clip,height=9cm]{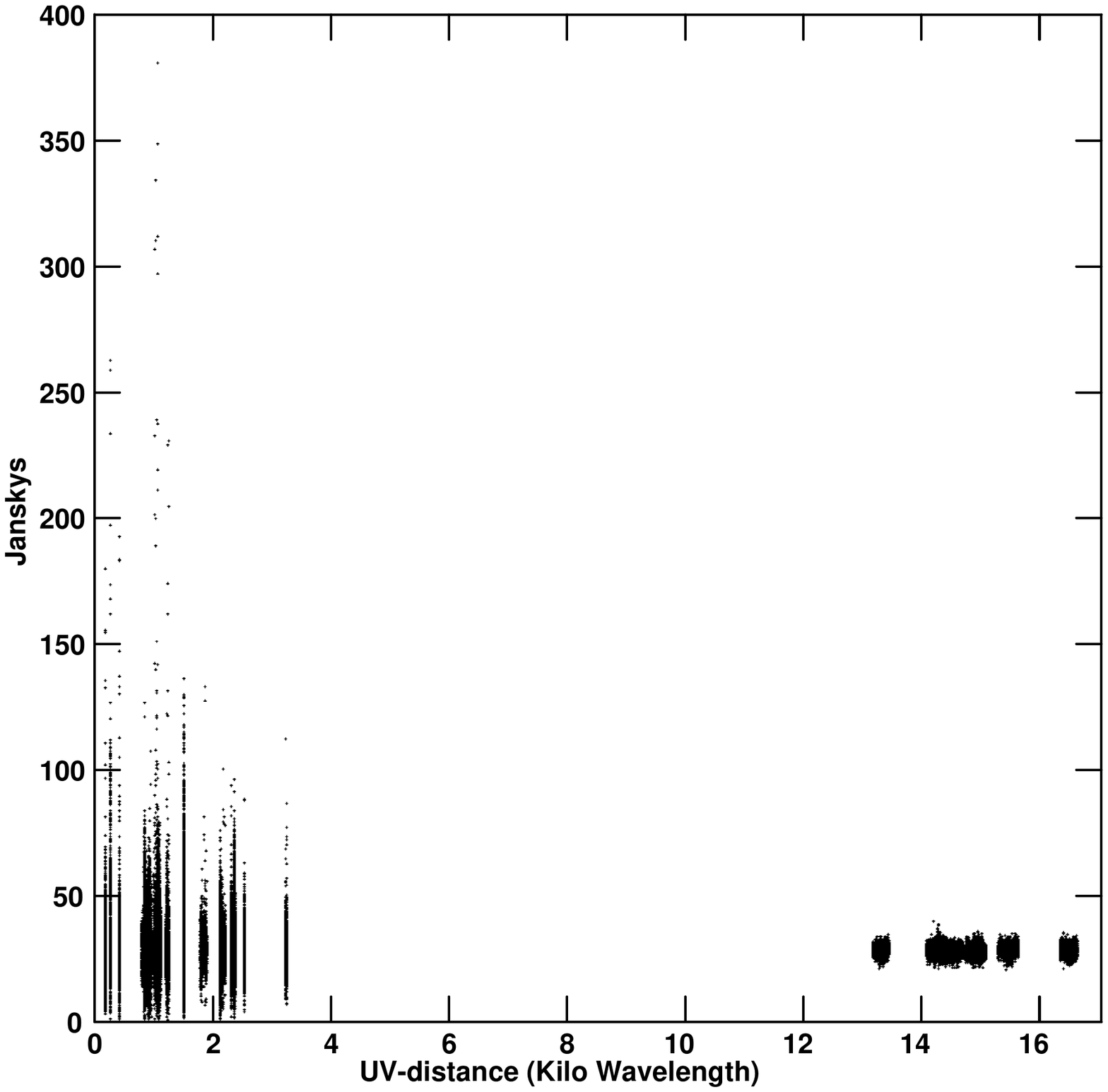}
        \includegraphics[trim={1cm 4cm 1cm 3cm}, clip,height=9cm]{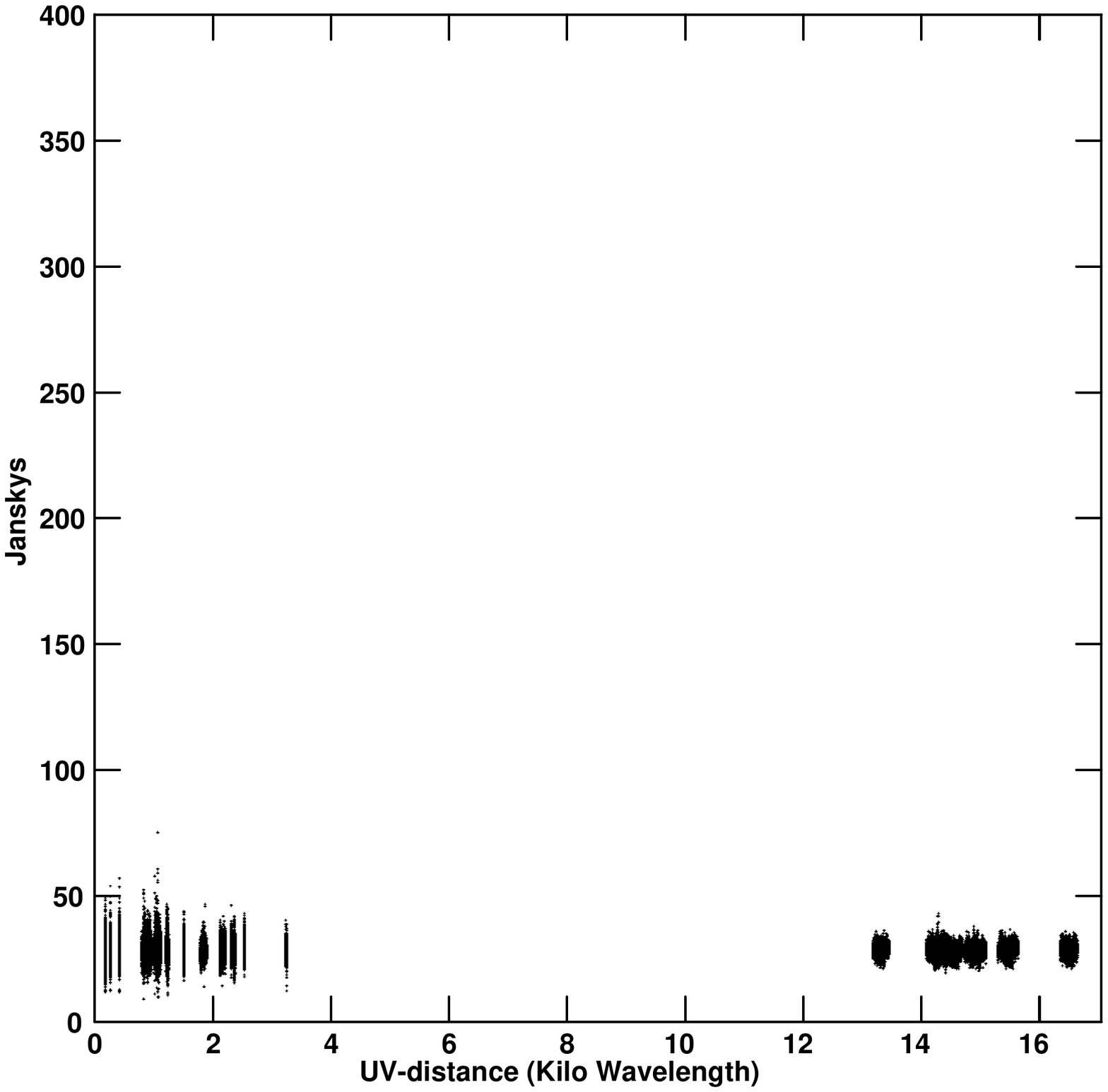}
 \caption{Calibrated amplitude versus UV-distance for unfiltered (left) and filtered data (right). A single spectral channel and both RR and LL polarizations are plotted. Short baselines where the broadband RFI is correlated (uv-distance $<4$ kilowavelengths) show reduction in the scatter around the mean amplitude.  
}
 \label{uvplt}
 \end{center}
\end{figure}

The distribution of calibrated visibility amplitudes around the mean value for different baseline lengths is a helpful test for understanding filtering. 
The visibilities from the recorded data are separated into unfiltered and filtered baselines. These test data are then processed separately using CASA (Common Astronomy and Software Applications, \citet{mcmullin2007casa})\footnote{https://casa.nrao.edu/} whilst using identical analysis steps. The flux density calibration was carried out using the standard flux density scale provided by \citet{perley2017accurate}.
The distribution of the calibrated visibility amplitudes (units in Jansky) of all timestamps in the test observation for different baseline lengths (in kilowavelengths) is shown in Fig. \ref{uvplt}. For an unresolved source observed in the absence of RFI, the values of visibility amplitude should remain identical across all the baseline lengths. The correlation of RFI on short baselines (uv-distance $<4$ kilowavelengths corresponding to about 2 km. from the centre of the array) results in a considerable variation for unfiltered data than those on longer baselines where the RFI is not correlated. For the filtered counterpart, the variations on both short and long baselines are comparable, indicating that the filter has removed the RFI.\\

\begin{figure}[h]
\begin{center}
        \includegraphics[trim={1cm 4cm 1cm 3cm}, clip,height=9cm]{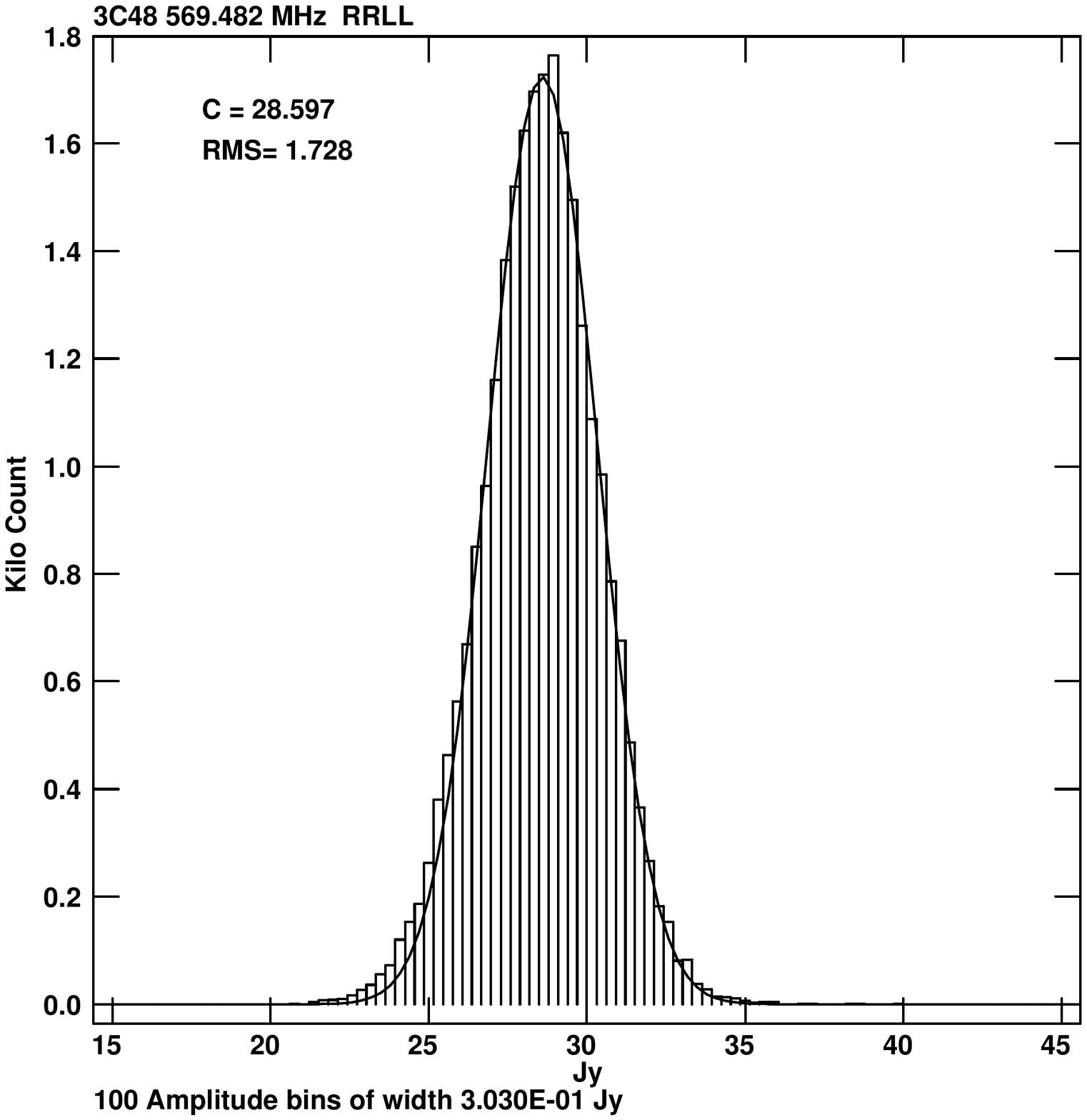}
        \includegraphics[trim={1cm 4cm 1cm 3cm}, clip,height=9cm]{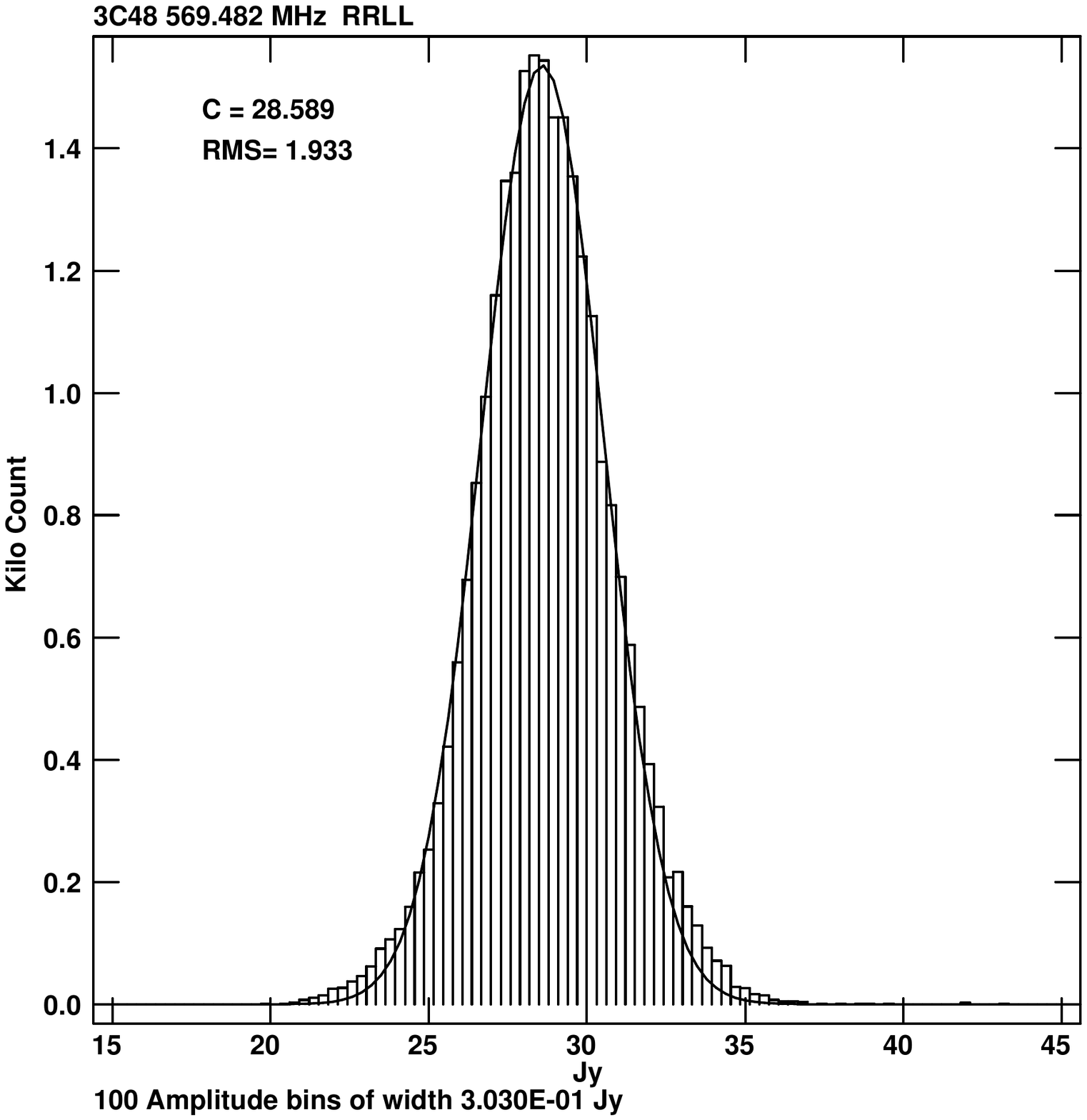}
 \caption{Histogram of calibrated amplitude for unfiltered (left) and filtered data (right) for the uv-distance $>4$ kilowavelengths. A single spectral channel and both RR and LL polarizations are plotted. There is an $11\%$ increase in the RMS in filtered amplitudes as compared to the unfiltered ones. The value of C gives the mean flux density in Jy of 3C48 in both the panels. 
}
 \label{uvhist}
 \end{center}
\end{figure}

On long baselines ($>4$ kilowavelengths), we noticed a slight increase in the scatter on the amplitudes around the mean value. This can be seen in Fig.~\ref{uvhist} where the histograms of the amplitudes for uv-distances $>4$ kilowavelengths for filtered and unfiltered data are plotted. This undesirable effect of the filtering was located and quantified using these plots. In this case, we see an increased scatter by $11\%$. On these long baselines, the replacement by digital noise results in introducing an uncorrelated component in the data resulting in this effect. 

\section{Discussion}
The techniques and data analysis methods described in this work were used to test the effect of real-time broadband RFI filtering on the beamformer and correlator data. It is not possible to record unfiltered and filtered data simultaneously for the entire array of 30 antennas. Thus, using the digital copy of signals, modes that allowed simultaneous recording of unfiltered and filtered data from half the array (1:2) were used. To simultaneously test more thresholds and replacement methods settings, the 1:2 mode was extended to allow a 1:4 digital copy. The recorded data were then analyzed offline, and the diagnostics used to assess the real-time RFI filtering were described. 

The CCF was used to quantify the effectiveness of the RFI filtering.
It is difficult to theoretically or empirically compute the mean CCF and error for all the baselines for a given observing run. Hence, the method adopted as part of the test procedure compares the CCF values of filtered data with the mean CCF and error bar for long baselines. Since powerline RFI is spatially correlated over shorter baselines, the CCF values from the long baselines can be used as a reference for comparing between unfiltered and filtered data.

Simultaneous testing using methods described here was successfully applied on various test observations to validate RFI filtering. For example, the 1:2 digital copy mode was useful in finding the effect of RFI filtering on long baselines in the cross-correlation (visibility) data. It was observed, through comparison between the unfiltered and filtered data, that the standard deviation on CCF was more significant for filtered data (Fig.~\ref{uvhist}). This effect has a dependence on the data loss and hence the filtering threshold. The exact reason for this effect on long baselines is being investigated.

Analyzing the effect of different replacement options and optimizing the filtering threshold is being carried out using the 1:4 digital copy mode described in the paper. This mode allows simultaneous testing and a comparative analysis between the unfiltered and filtered data with different threshold values and replacement options. For example, simultaneous testing of the filtering technique at 3, 4, and 5$\sigma$ thresholds and digital noise, zero (blanking), and threshold (clipping) are being carried out for test observations. A variant of the 1:4 design has been developed to provide specific antenna combinations that are useful in studying the effect of filtering on longer baselines.

\section{Conclusion} 
The paper presented techniques developed for evaluating the performance of the real-time broadband RFI filtering system at the uGMRT. An example uGMRT test observation was considered as an illustration. Performance analysis methods applicable to beamformer and correlator outputs were introduced and explained. The concept of simultaneous testing and its significance in understanding the performance of the filter was highlighted. The effects of filtering on astronomical data and fine-tuning the filtering parameters was described. These techniques can be extended for evaluating the performance of any real-time or offline broadband or narrowband RFI mitigation technique in array radio telescopes. Further, it can also be used for identification, monitoring, and understanding the effects of RFI on telescope receiver systems and astronomical data.

\section{Acknowledgments}

We would like to thank the past team members who worked on the real-time RFI filtering system. We acknowledge the GMRT control room, Operations group, and Backend group members for their help in carrying out the RFI filtering test observations. We are thankful to Pravin Raybole and RFI group at GMRT for discussions on potential sources of powerline RFI and their locations. We would also like to express our gratitude towards NCRA faculty members for their suggestions and feedback on the filtering system.

\bibliographystyle{ws-jai}
\bibliography{jai_intro}

\clearpage
\end{document}